\documentclass[epj]{svjour}

\usepackage{graphicx}
\usepackage{amssymb}

\begin{document}

\title{Two-dimensional $t$-$J$ model at moderate doping}

\author{A.~Sherman\inst{1} \and M.~Schreiber\inst{2}}

\institute{Institute of Physics, University of Tartu, Riia 142, 51014
Tartu, Estonia \and Institut f\"ur Physik, Technische Universit\"at,
D-09107 Chemnitz, Federal Republic of Germany\\
and School of Engineering and Science, International University Bremen,
Campus Ring 1, D-28759 Bremen,
\\ Federal Republic of Germany}

\date{Received: date / Revised version: date}

\abstract{Using the method which retains the rotation symmetry of spin
components in the paramagnetic state and has no preset magnetic
ordering, spectral and magnetic properties of the two-dimensional
$t$-$J$ model in the normal state are investigated for the ranges of
hole concentrations $0 \leq x \leq 0.16$ and temperatures $0.01t \leq T
\leq 0.2t$. The used hopping $t$ and exchange $J$ parameters of the
model correspond to hole-doped cuprates. The obtained solutions are
homogeneous which indicates that stripes and other types of phase
separation are not connected with the strong electron correlations
described by the model. A series of nearly equidistant maxima in the
hole spectral function calculated for low $T$ and $x$ is connected with
hole vibrations in the region of the perturbed short-range
antiferromagnetic order. The hole spectrum has a pseudogap in the
vicinity of $(0,\pi)$ and $(\pi,0)$. For $x \approx 0.05$ the shape of
the hole Fermi surface is transformed from four small ellipses around
$(\pm\pi/2,\pm\pi/2)$ to two large rhombuses centered at $(0,0)$ and
$(\pi,\pi)$. The calculated temperature and concentration dependencies
of the spin correlation length and the magnetic susceptibility are
close to those observed in cuprate perovskites. These results offer
explanations for the observed scaling of the static uniform
susceptibility and for the changes in the spin-lattice relaxation and
spin-echo decay rates in terms of the temperature and doping variations
in the spin excitation spectrum of the model.
\PACS{
      {71.10.Fd}{Lattice fermion models}   \and
      {74.25.Ha}{Magnetic properties} \and
      {74.25.Jb}{Electronic structure}
     } 
} 

\maketitle

\section{Introduction}
The two-dimensional $t$-$J$ model was proposed by Anderson
\cite{Anderson} for the description of strong electron correlations in
CuO$_2$ planes of perovskite high-$T_c$ superconductors. In
Ref.~\cite{FZhang} the similarity of the low-energy part of its
spectrum with the spectrum of the realistic three-band Hubbard model
was demonstrated. Nowadays the $t$-$J$ model is one of the most
frequently used models for the interpretation of experimental results
in cuprates (for a review, see Ref.~\cite{Izyumov}). Different
numerical and analytical methods were used for the investigation of the
model. Among these methods are the exact diagonalization of small
clusters \cite{Dagotto,Jaklic}, Monte Carlo simulations \cite{SZhang},
density matrix renormalization group calculations \cite{White},
spin-wave \cite{Sherman98,Plakida} and mean-field slave-boson
approximations \cite{Kane}. In spite of the considerable progress made
towards the understanding of the properties of the model, the basic
issues of its behavior at moderate doping have not yet been completely
resolved.

Aiming at a description for this range of concentration in
Ref.~\cite{Sherman02} a new analytical method was developed which has
merits of retaining the rotation symmetry of spin components in the
paramagnetic state and of the absence of any preset magnetic ordering.
Since the description is carried out in terms of the Hubbard operators,
the method takes proper account of the kinematic interaction. The
method is based on Mori's projection operator technique \cite{Mori}
which allows one to represent Green's functions in the form of
continued fractions and gives a way for calculating their elements. The
residual term of the fraction is approximated by the decoupling which
reduces this many-particle Green's function to a product of the
sought-for functions. Using the idea of Ref.~\cite{Kondo} the
decoupling is corrected by introducing a vertex correction, a
multiplier which is determined from the sum rule of the considered
problem, the constraint of zero site magnetization in the paramagnetic
state. Test calculations with this method for small clusters and high
temperatures demonstrated good agreement of the obtained results with
the exact diagonalization and Monte Carlo data. Notice that close
approaches which use equations of motion and Tserkovnikov's formalism
were developed in Ref.~\cite{Shimahara91} to consider magnetic
properties of the 2D $t$-$J$ model.

This article contains results of calculations carried out with the
method of Ref.~\cite{Sherman02} for a 20$\times$20 lattice and
parameters $J=0.1$~eV, $t=0.5$~eV where $J$ and $t$ are the exchange
and hopping constants of the $t$-$J$ model. Test calculations with
lattices of other sizes showed that the considered lattice is large
enough to avoid finite-size effects in spectral functions. The
parameters chosen correspond to hole-doped cuprate perovskites
\cite{Mcmahan}. The considered range of hole concentrations $0 \leq x
\leq 0.16$ spans the cases from light to moderate doping. The
temperature range of the calculations $0.01t \approx 58\,{\rm K} \leq T
\leq 0.2t \approx 1200\,{\rm K}$ embraces the range of temperatures
used in experiment.

The calculated Green's functions reveal a number of peculiarities in
the hole and spin excitation spectrum. For low $x$ and $T$, besides the
spin-polaron peak \cite{Izyumov}, a series of less intensive and
broader maxima is observed at higher frequencies in the hole spectral
function. These maxima are nearly equidistant which allows us to
connect them with vibrations of a hole in a region of the perturbed
short-range antiferromagnetic order. This region arises due to the hole
movement. The lowest of these maxima with a dispersion which mimics the
dispersion of the spin-polaron peak is apparently observed in lightly
doped Ca$_2$CuO$_2$Cl$_2$. For low hole concentrations only the
spin-polaron band crosses the Fermi level which leads to the Fermi
surface consisting of four ellipses around $(\pm\pi/2,\pm\pi/2)$. For
$x \approx 0.05$ the second band, which arises below the Fermi level in
the used hole picture, crosses the Fermi level and the Fermi surface
acquires new elements -- large rhombuses around $(0,0)$ and
$(\pi,\pi)$. However, due to the peculiar dispersion of the
spin-polaron band and its larger spectral intensity parts of this large
Fermi surface near $(0,\pi)$ and $(\pi,0)$ are hidden which looks like
a pseudogap, a decrease of the low-frequency spectral intensity in
these regions of the Brillouin zone. The magnitude of the pseudogap --
the energy distance between the spin-polaron peak and the Fermi level
-- decreases with $x$ and at $x \approx 0.12$ the pseudogap disappears.
The symmetry, magnitude and concentration dependence of the calculated
pseudogap are similar to those observed in Bi-based cuprates.

As mentioned, the method used has no preset magnetic ordering which
opens the way to investigate whether the strong electron correlations
described by the $t$-$J$ model are responsible for the charge and
magnetic inhomogeneities observed in some cuprates \cite{Tranquada}. In
the considered ranges of hole concentrations and temperatures only the
homogeneous solutions were found which indicates that other
interactions have to be invoked to explain these inhomogeneities.

The calculated spectrum of spin excitations contains a gap near
$(\pi,\pi)$ which is directly connected with the correlation length of
the short-range antiferromagnetic order. The dependence of the
correlation length on $x$ reproduces the relation observed
experimentally in La$_{2-x}$Sr$_x$CuO$_4$ \cite{Keimer}. With
increasing $x$ the branch of spin excitations is destroyed near
$(0,0)$, whereas at the periphery of the Brillouin zone the excitations
remain well defined. The calculated magnetic susceptibility and its
dependencies on $x$ and $T$ are similar to those observed in neutron
scattering and NMR experiments. This similarity allows us to offer
explanations for the observed scaling of the static uniform
susceptibility and for the changes in the spin-lattice relaxation and
spin-echo decay rates in terms of the temperature and doping variations
in the spin excitation spectrum of the $t$-$J$ model.

For convenience main formulas of Ref.~\cite{Sherman02}, which were used
in the calculations, are reproduced in Sec.~II\@. In Sec.~III and IV
peculiarities of the hole and spin excitation spectra are discussed.
The spin susceptibility, spin correlations, spin-lattice relaxation and
spin-echo decay rates are considered in Sec.~V\@. Our conclusions are
given in Sec.~VI.

\section{Main formulas}
The Hamiltonian of the 2D $t$-$J$ model reads \cite{Izyumov}
\begin{equation}
H=\sum_{\bf nm\sigma}t_{\bf nm}a^\dagger_{\bf n\sigma}a_{\bf
m\sigma}+\frac{1}{2}\sum_{\bf nm}J_{\bf nm}\left(s^z_{\bf n}s^z_{\bf
m}+s^{+1}_{\bf n}s^{-1}_{\bf m}\right), \label{hamiltonian}
\end{equation}
where $a_{\bf n\sigma}=|{\bf n}\sigma\rangle\langle{\bf n}0|$ is the
hole annihilation operator, {\bf n} and {\bf m} label sites of the
square lattice, $\sigma=\pm 1$ is the spin projection, $|{\bf
n}\sigma\rangle$ and $|{\bf n}0\rangle$ are site states corresponding
to the absence and presence of a hole on the site. These states may be
considered as linear combinations of the products of the $3d_{x^2-y^2}$
copper and $2p_\sigma$ oxygen orbitals of the extended Hubbard model
\cite{Sherman98,Jefferson}. In this work we take into account nearest
neighbor interactions only, $t_{\bf nm}=-t\sum_{\bf a}\delta_{\bf
n,m+a}$ and $J_{\bf nm}=J\sum_{\bf a}\delta_{\bf n,m+a}$ where the four
vectors {\bf a} connect nearest neighbor sites. The spin-$\frac{1}{2}$
operators can be written as $s^z_{\bf
n}=\frac{1}{2}\sum_\sigma\sigma|{\bf n}\sigma\rangle\langle{\bf
n}\sigma|$ and $s^\sigma_{\bf n}=|{\bf n}\sigma\rangle\langle{\bf
n},-\sigma|$.

Due to the complicated commutation relations the diagram technique for
the operators $a_{\bf n\sigma}$, $s^z_{\bf n}$, and $s^\sigma_{\bf n}$
is very intricate \cite{Skryabin}. In this case the use of Mori's
projection operator technique \cite{Mori} for the derivation of
self-energy equations for Green's functions constructed from such
operators is especially fruitful. In this way the following equations
were found for the hole $G({\bf k}t)=\langle\!\langle a_{\bf
k\sigma}|a_{\bf k\sigma}^\dagger\rangle\!\rangle
=-i\theta(t)\langle\!\{a_{\bf k\sigma}(t),a^\dagger_{\bf
k\sigma}\}\!\rangle$ and spin $D({\bf k}t)=-i\theta(t)\langle[s^z_{\bf
k}(t), s^z_{\bf -k}]\rangle$ Green's functions \cite{Sherman02}:
\begin{eqnarray}
D({\bf k}\omega)&=&\frac{[4J\alpha(\Delta+1+\gamma_{\bf
k})]^{-1}\Pi({\bf k}\omega)+4JC_1(\gamma_{\bf k}-1)}{\omega^2-\Pi({\bf
k}\omega)- \omega^2_{\bf k}}, \nonumber\\
&&\label{se} \\
G({\bf k}\omega)&=&\phi[\omega-\varepsilon_{\bf k}+\mu-\Sigma({\bf
k}\omega)]^{-1}, \nonumber
\end{eqnarray}
where $D({\bf k}\omega)=\int^\infty_{-\infty}\exp(i\omega t)D({\bf
k}t)dt$,
\begin{eqnarray*}
a_{\bf k\sigma}&=&N^{-1/2}\sum_{\bf n}\exp(-i{\bf kn})a_{\bf
n\sigma},\\
s^z_{\bf k}&=&N^{-1/2}\sum_{\bf n}\exp(-i{\bf kn})s^z_{\bf n},
\end{eqnarray*}
$a_{\bf k\sigma}(t)=\exp(i{\cal H}t)a_{\bf k\sigma}\exp(-i{\cal H}t)$,
$N$ is the number of sites, ${\cal H}=H-\mu\sum_{\bf n}X_{\bf n}$,
$\mu$ is the chemical potential, $X_{\bf n}=|{\bf n}0\rangle
\langle{\bf n}0|$, the angular brackets denote averaging over the grand
canonical ensemble, $\gamma_{\bf k}=\frac{1}{4}\sum_{\bf a}\exp(i{\bf
ka})$, $\phi=\frac{1}{2}(1+x)$, and
\begin{eqnarray}
\omega^2_{\bf k}&=&16J^2\alpha|C_1|(1-\gamma_{\bf k})
(\Delta+1+\gamma_{\bf k}),\nonumber\\
&&\label{seed}\\
\varepsilon_{\bf k}&=&-(4\phi
t+6C_1\phi^{-1}t+3F_1\phi^{-1}J)\gamma_{\bf k}.\nonumber
\end{eqnarray}

In the present calculations the parameter of vertex correction
$\alpha$, which improves the decoupling in the residual term, is set
equal to its value in the undoped case, $\alpha=1.802$ [this value
differs slightly from that reported in Refs.~\cite{Sherman02,Shimahara}
due to an artificial broadening introduced in $D({\bf k}\omega)$, see
below]. The parameter $\Delta$ which describes a gap in the spectrum of
spin excitations at $(\pi,\pi)$ [see Eq.~(\ref{seed})] is determined by
the constraint of zero site magnetization $\langle s^z_{\bf
l}\rangle=0$ in the paramagnetic state. The constraint can be written
in the form
\begin{equation}\label{zsm}
\frac{1}{2}(1-x)=\frac{2}{N}\sum_{\bf k}\int_0^\infty d\omega
\coth\!\left(\frac{\omega}{2T}\right)B({\bf k}\omega),
\end{equation}
where $B({\bf k}\omega)= -\pi^{-1}{\rm Im}\,D({\bf k}\omega)$ is the
spin spectral function. Notice that in accord with the Mermin-Wagner
theorem \cite{Mermin} in the considered 2D system the long-range
antiferromagnetic ordering is destroyed at any nonzero $T$ and, as will
be seen below, for $T=0$ and $x > x_c \approx 0.02$. The value of $x$
and the nearest neighbor correlations $C_1=\langle s_{\bf l}^{+1}s_{\bf
l+a}^{-1}\rangle$ and $F_1=\langle a_{\bf l}^\dagger a_{\bf
l+a}\rangle$ are connected with Green's functions (\ref{se}) by the
following relations:
\begin{eqnarray}
x&=&\frac{1}{N}\sum_{\bf k}\int_{-\infty}^\infty d\omega
 n_F(\omega)A({\bf k}\omega),\nonumber\\
F_1&=&\frac{1}{N}\sum_{\bf k}\gamma_{\bf k}\int_{-\infty}^\infty
 d\omega n_F(\omega)A({\bf k}\omega),\label{xcf}\\
C_1&=&\frac{2}{N}\sum_{\bf k}\gamma_{\bf k}\int_0^\infty d\omega
 \coth\!\left(\frac{\omega}{2T}\right)B({\bf k}\omega),\nonumber
\end{eqnarray}
where $A({\bf k}\omega)=-\pi^{-1}{\rm Im}\,G({\bf k}\omega)$ is the
hole spectral functions and $n_F(\omega)=[\exp(\omega/T)+1]^{-1}$.

The self-energies in Eq.~(\ref{se}) read
\begin{eqnarray}
{\rm Im}\,\Pi({\bf k}\omega)&=&\frac{16\pi t^2J}{N}(\Delta+1+
 \gamma_{\bf k})\sum_{\bf k'}(\gamma_{\bf k}-\gamma_{\bf
 k+k'})^2 \nonumber\\
&\times&\int^\infty_{-\infty}d\omega'[n_F(\omega+\omega')-
 n_F(\omega')] \nonumber\\
&\times&A({\bf k+k'},\omega+\omega')A({\bf k'}\omega'), \nonumber\\
{\rm Im}\,\Sigma({\bf k}\omega)&=&\frac{16\pi t^2}{N\phi}\sum_{\bf k'}
 \int_{-\infty}^\infty d\omega'\biggl[\gamma_{\bf k-k'}+\gamma_{\bf
 k} \label{po}\\
&+&{\rm sgn}(\omega')(\gamma_{\bf
 k-k'}-\gamma_{\bf k})\sqrt{\frac{1+\gamma_{\bf k'}}{1-\gamma_{\bf
 k'}}}\biggr]^2 \nonumber\\
&\times&[n_B(-\omega')+n_F(\omega-\omega')] \nonumber\\
&\times&A({\bf k-k'},\omega-\omega')B({\bf k'}\omega'), \nonumber
\end{eqnarray}
where $n_B(\omega)=\left[ \exp(\omega/T)-1\right]^{-1}$. The source of
damping of spin excitations described by Eq.~(\ref{po}) is the decay
into two fermions. Another source of damping, multiple spin excitation
scattering, is considered phenomenologically by adding the small
artificial broadening $-2\eta\omega_{\bf k}$, $\eta=0.02t$ to ${\rm
Im}\Pi({\bf k}\omega)$. The broadening $-\eta$ is also added to ${\rm
Im}\Sigma({\bf k}\omega)$ to widen narrow lines and to stabilize the
iteration procedure.

The same derivation for the transversal spin Green's function gives
$\langle\!\langle s_{\bf k}^{-1}\big|s_{\bf k}^{+1}\rangle\!
\rangle=2D({\bf k}t)$ indicating that the used approach retains
properly the rotation symmetry of spin components in the paramagnetic
state.

For low hole concentrations and temperatures the bandwidth of the
dispersion $\varepsilon_{\bf k}$, Eq.~(\ref{seed}), is approximately
equal to $t$ which is much smaller than $8t$, the bandwidth of
uncorrelated electrons. The reason for this band narrowing is the
antiferromagnetic alignment of spins. In this case the hole movement is
accompanied by the spin flipping. In these conditions the hole
dispersion is determined by the self-energy $\Sigma({\bf k}\omega)$
which is responsible for a considerable energy gain for states in the
spin-polaron band. For the parameters used this gain is approximately
equal to $2t$.

\section{The hole spectrum}
\begin{figure}
\centerline{\includegraphics{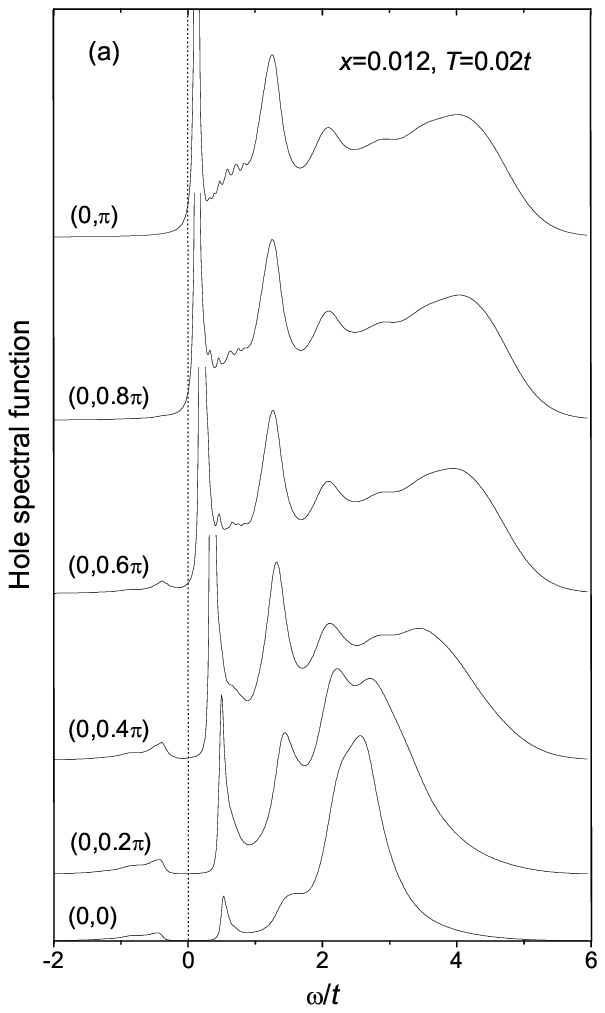}}

\vspace{4mm}\centerline{\includegraphics{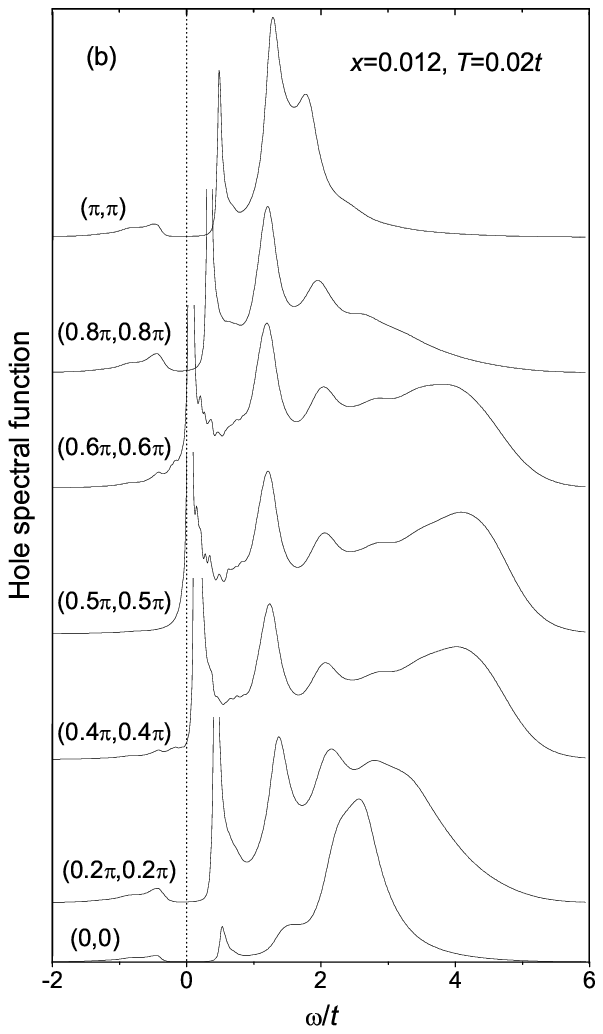}}
\caption{\label{Fig_i}The hole spectral function $A({\bf k}\omega)$
along the symmetry lines for $x=0.012$, $T=0.02t$ and $J/t=0.2$. The
respective values of the wave vectors are indicated near the curves.
Here and below the hole picture is used. The vertical dotted line
indicates the position of the chemical potential.}
\end{figure}

\begin{figure}[!]
\centerline{\includegraphics{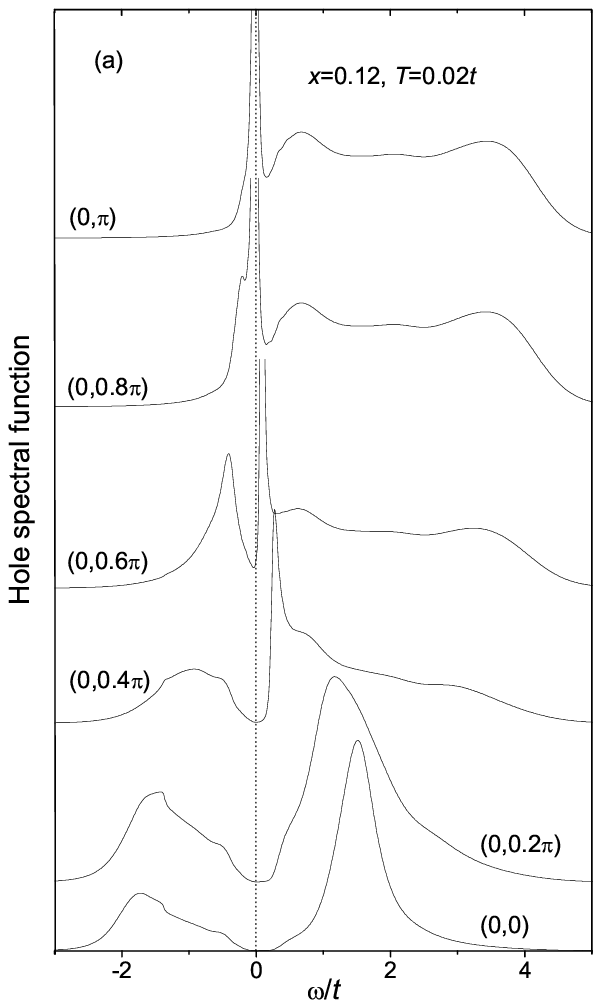}}

\vspace{4mm}\centerline{\includegraphics{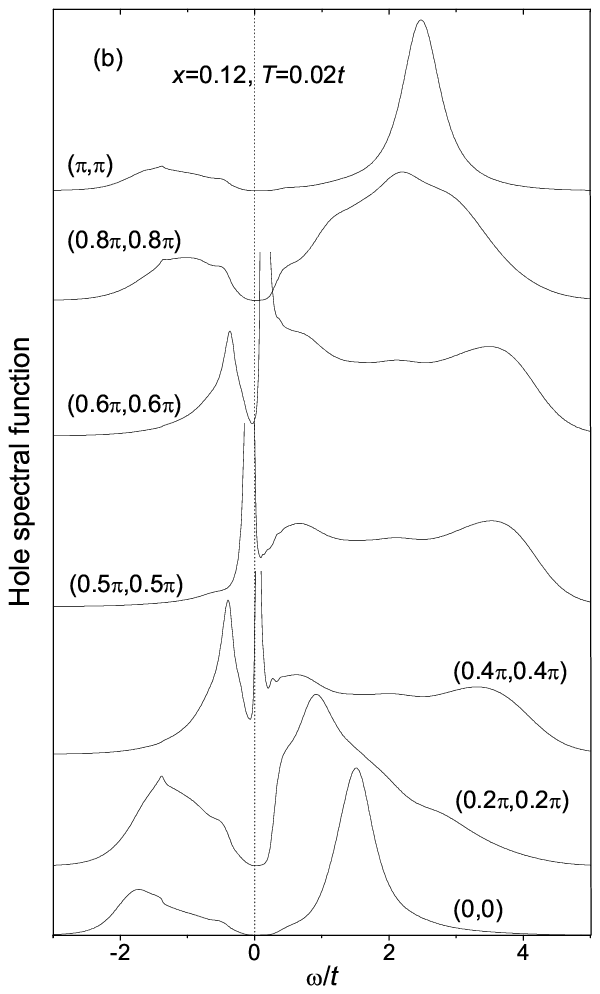}}
\caption{\label{Fig_ii}The hole spectral function for $x=0.12$ and
$T=0.02t$.}
\end{figure}

\begin{figure}
\centerline{\includegraphics[width=7.8cm]{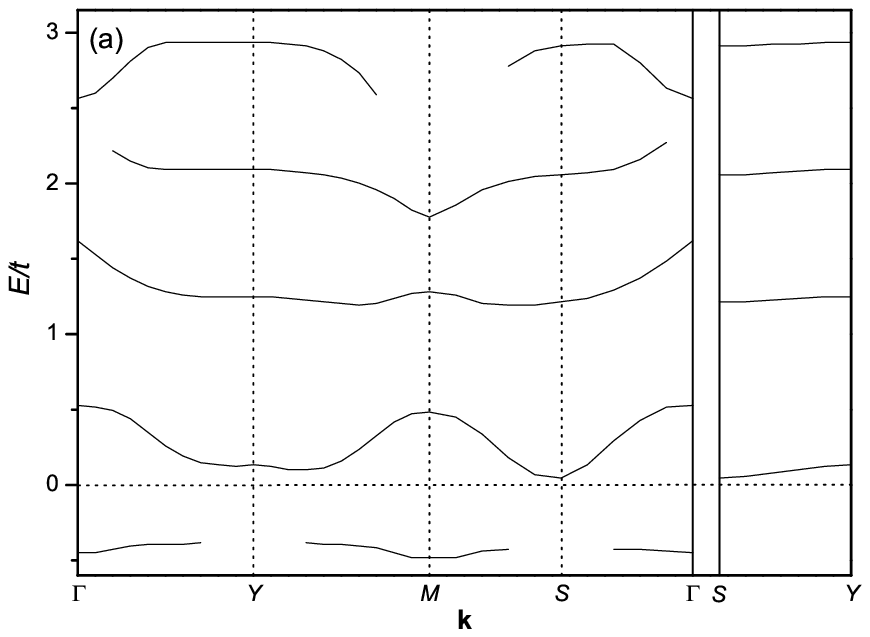}}

\vspace{2mm}\centerline{\includegraphics[width=8.0cm]{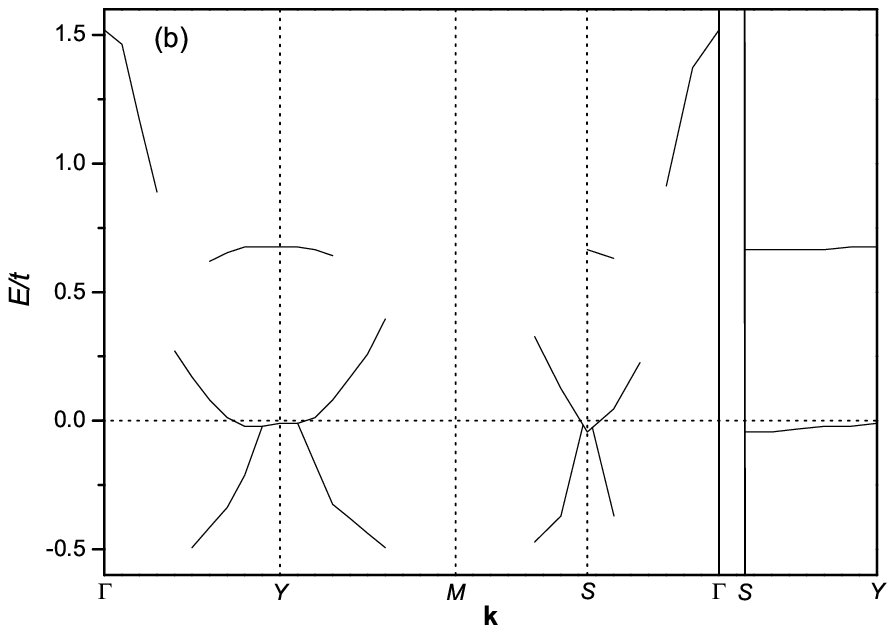}}
\caption{\label{Fig_iii}The dispersion of the maxima in the hole
spectral function for $T=0.02t$, $x=0.012$ (a) and $x=0.12$ (b). Points
$Y$, $M$ and $S$ correspond to ${\bf k}=(0,\pi)$, $(\pi,\pi)$ and
$(\pi/2,\pi/2)$, respectively.}
\end{figure}

Equations~(\ref{se})--(\ref{po}) form a closed set which was solved by
iteration. Examples of the hole spectral functions for the cases of low
and moderate doping are given in Figs.~\ref{Fig_i} and \ref{Fig_ii}.%
The frequency is measured from the chemical potential. For low $x$ the
shapes of the spectra are close to those obtained in the spin-wave
approximation \cite{Sherman98}. As seen from Fig.~\ref{Fig_i}, at low
doping the hole spectral function contains a series of nearly
equidistant maxima. The first member of this series is the narrow peak
with the highest intensity which is frequently termed the spin-polaron
peak (in some of the spectra in Figs.~\ref{Fig_i} and \ref{Fig_ii} only
its foot is shown). The dispersion of the maxima in the above spectra
along the symmetry directions is shown in Fig.~\ref{Fig_iii}. In the
spin-polaron band the lowest energy is reached at $(\pi/2,\pi/2)$.
The nearly equidistant location of the maxima is clearly seen in
Fig.~\ref{Fig_iii}a. Besides, we notice that the second maximum in
the series has a dispersion which is close in shape to the
dispersion of the spin-polaron peak. As follows from
Figs.~\ref{Fig_ii} and \ref{Fig_iii}b, this maximum is retained at
moderate doping in some region of the Brillouin zone.

The nearly equidistant position of the maxima in the low-concentration
spectra allows us to connect them with vibronic states of a hole in the
region of the perturbed short-range antiferromagnetic order. The moving
hole le\-a\-ves behind it a trace of overturned spins. Such disturbance
requires energy which in its turn leads to a restoring force acting on
the hole and giving rise to its vibrations. This notion is close to the
string picture developed in Ref.~\cite{Siggia} for the Ising model and
used for the interpretation of the fine spectral structure in the
$t$-$J$ model on small lattices \cite{Izyumov,Dagotto}.

The two lowest maxima of the above-mentioned series are apparently
observed in lightly doped Ca$_2$CuO$_2$Cl$_2$ \cite{Kim}. The narrow
and intensive spin-polaron peak is responsible for the photoemission
maximum with the lowest binding energy, while the second member of the
series corresponds to the spectral maximum which is observed at
approximately 600~meV higher binding energy. This energy difference is
close to that shown in Fig.~\ref{Fig_iii}a. Besides, as noted in
Ref.~\cite{Kim}, the dispersion of the second maximum mimics the
dispersion of the spin-polaron peak, which conforms with our
calculations. However, it should be noted that our calculated
dispersion of the spin-polaron band differs somewhat from that observed
experimentally in lightly doped cuprates \cite{Kim,Wells}. Although the
shapes and magnitudes of the experimental and calculated dispersions
along the nodal direction [from $(0,0)$ to $(\pi,\pi)$] are close,
along the boundary of the magnetic Brillouin zone [from $(0,\pi)$ to
$(\pi,0)$] the magnitude of the calculated dispersion is an order of
magnitude smaller than the experimental one. A possible reason for this
discrepancy is the oversimplified hole hopping term in Hamiltonian
(\ref{hamiltonian}) which takes into account the transfer between
nearest neighbor sites only \cite{Kim,Barabanov}.

The spectral intensity below the chemical potential in
Figs.~\ref{Fig_i} and \ref{Fig_iii}a is noteworthy. Already at low $x$
in some region of the Brillouin zone this spectral feature looks like a
weak maximum. It is this spectral intensity which provides a finite
hole concentration until the spin-polaron band crosses the Fermi level
at $x \approx 0.04$.

With increasing $x$ and $T$ the high-frequency maxima in the
nearly equidistant series are smeared, as seen in
Fig.~\ref{Fig_ii}. Inspecting this figure one can see that for
moderate hole concentrations two maxima -- the spin-polaron peak
and a less intensive and broader maximum -- cross the Fermi level.
This latter maximum originates from the mentioned spectral feature
which appears below the chemical potential at low $x$. The
dispersions of the two maxima and some other spectral
peculiarities are shown in Fig.~\ref{Fig_iii}b. The maxima can be
resolved in the spectrum only in some regions of the Brillouin
zone. This is the reason why the curves in Fig.~\ref{Fig_iii}b
terminate at some points of the symmetry lines.

\begin{figure}
\centerline{\includegraphics[width=4.1cm]{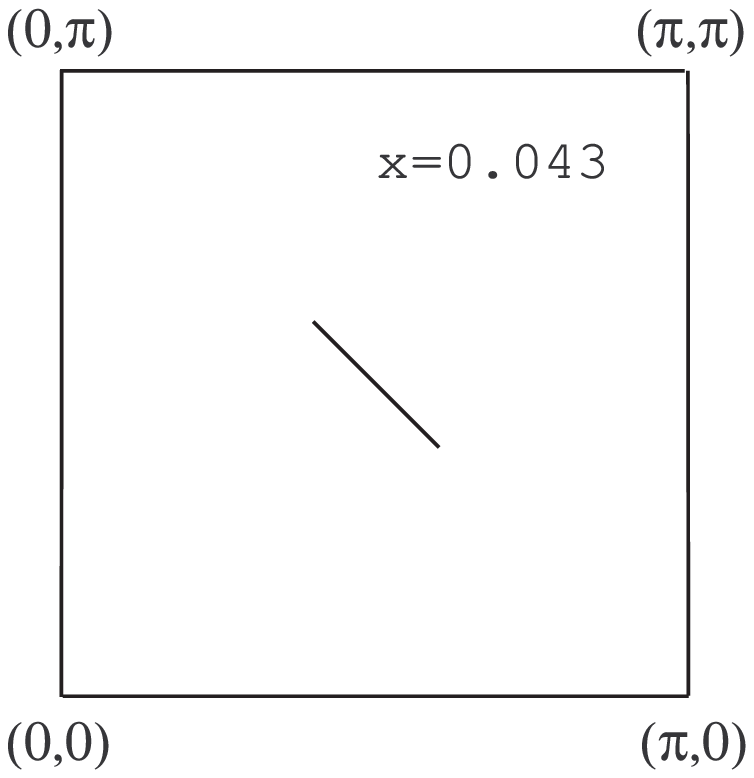}}

\vspace{2mm}\centerline{\includegraphics[width=3.5cm]{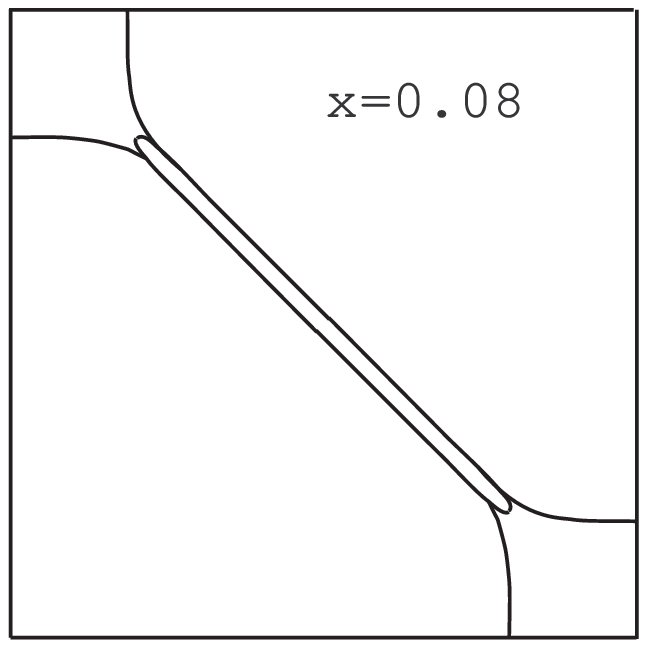}}

\vspace{2mm}\centerline{\includegraphics[width=3.5cm]{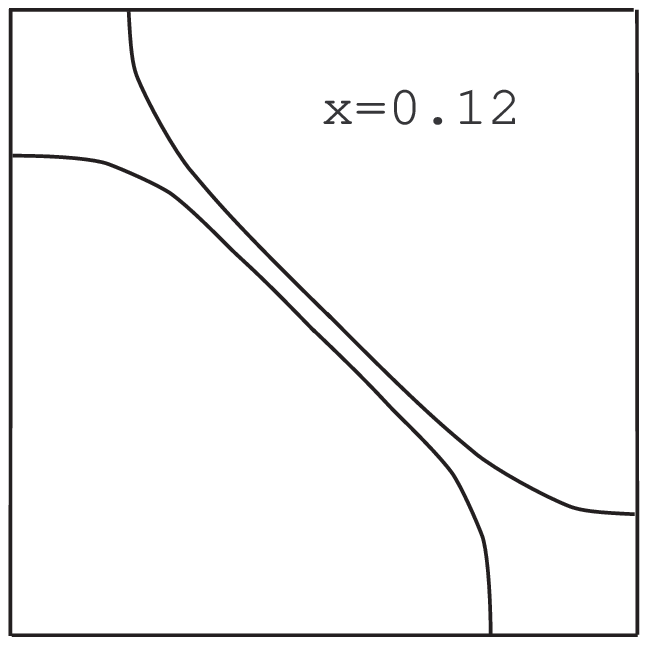}}
\caption{\label{Fig_iv}The Fermi surface in the first quadrant of the
Brillouin zone for $x=0.043$, $0.08$, and $0.12$. $T=0.02t$.}
\end{figure}

The evolution of the Fermi surface with doping is shown in
Fig.~\ref{Fig_iv}.
No Fermi surface exists for $x \lesssim 0.04$, since no maximum which
can be identified with a quasiparticle excitation crosses the Fermi
level (see Figs.~\ref{Fig_i} and \ref{Fig_iii}a). As indicated above,
for such $x$ the occupied hole states are located at some distance from
the level. At $x \approx 0.04$ the spin-polaron band crosses the Fermi
level near $(\pm\pi/2,\pm\pi/2)$ which produces a Fermi surface
consisting of four strongly elongated ellipses around these points. In
Fig.~\ref{Fig_iv} the transverse size of one of these ellipses is much
smaller than the used momentum step and therefore the ellipse is
depicted as a line segment. At $x \approx 0.05$ the second band crosses
the Fermi level. With this crossing the Fermi surface acquires new
elements -- in addition to the ellipses of the spin-polaron band there
appear two large rhombuses centered at $(0,0)$ and $(\pi,\pi)$ (see
Fig.~\ref{Fig_iv}, the case $x=0.08$). One of these rhombuses can be
considered as a shadow image of the other, which arises due to the
antiferromagnetic short-range order. Notice, however, that the
rhombuses are slightly different in size. In such a manner the small
Fermi surface which exists for $x \lesssim 0.05$ is transformed to the
large surface for larger hole concentrations. With a further increase
of $x$ the Fermi level moves up from the bottom of the spin-polaron
band, the size of the ellipses grows and their shape is somewhat
changed. At $x \approx 0.12$ they reach the boundaries of the Brillouin
zone and the parts of the Fermi surface connected with the spin-polaron
band are transformed into two rhombuses the shapes and sizes of which
are close to those connected with the second band. Thus, for $x \gtrsim
0.12$ the Fermi surface consists of two rhombuses centered at $(0,0)$
and $(\pi,\pi)$ (see Fig.~\ref{Fig_iv}).

The compound Bi$_2$Sr$_2$CaCu$_2$O$_{8+\delta}$ is most intensively
investigated in angle-resolved photoemission experiments. However, even
for this crystal the Fermi surface topology is a topic of an intense
controversy (see, e.g., Ref.~\cite{Damascelli} and references therein).
The main reasons for this are broad spectral features in underdoped
normal-state crystals and the weak dispersion around $(\pi,0)$ (the so
called flat bands or extended van Hove singularities, see
Fig.~\ref{Fig_iii}b). We notice also that for moderate doping the
spin-polaron band in all the area between the rhombuses lies less than
20~meV below the Fermi level. This energy is on the verge of accuracy
of photoemission experiments. Nevertheless in the variety of
experimental results on the Fermi surface of normal-state cuprates it
is possible to find some features which resemble those shown in
Fig.~\ref{Fig_iv}. The shadow Fermi surfaces (together with some other
replicas of the main Fermi surface) are observed in
Bi$_2$Sr$_2$CaCu$_2$O$_{8+\delta}$ \cite{Damascelli,Aebi}. In some
experimental conditions in this crystal and in La$_{2-x}$Sr$_x$CuO$_4$
the shapes of experimental Fermi surfaces \cite{King,Zhou} are close to
that shown in Fig.~\ref{Fig_iv} for $x=0.12$. Notice however that in
Ref.~\cite{Zhou} such shape of the Fermi surface in
La$_{2-x}$Sr$_x$CuO$_4$ is ascribed to dynamic stripes presumed in this
crystal.

As indicated, in the range $0.05 \lesssim x \lesssim 0.12$ the Fermi
surface contains parts arising due to the crossings of the Fermi level
by two different bands. The spectral maxima corresponding to these
bands differ essentially in their intensity. In comparison with the
spin-polaron peak the spectral maximum of the second band is weaker and
bro\-a\-der (see Fig.~\ref{Fig_ii}). As a consequence, for wave vectors
near $(0,\pi)$ and $(\pi,0)$ this maximum is lost to the foot of the
spin-polaron peak on crossing the Fermi level. Since the spin-polaron
band lies somewhat above the Fermi level in this region of the
Brillouin zone, the situation looks like a part of the Fermi surface
disappears here and a gap opens between the hole energy band and the
Fermi level \cite{Sherman98,Sherman97}. In its size ($\sim 20$~meV) and
symmetry this gap is similar to the pseudogap observed in photoemission
spectra. The gap disappears at $x \approx 0.12$ when the whole Fermi
surface stems from the crossing of the spin-polaron band with the Fermi
level.

\begin{figure}[t]
\centerline{\includegraphics[width=8cm]{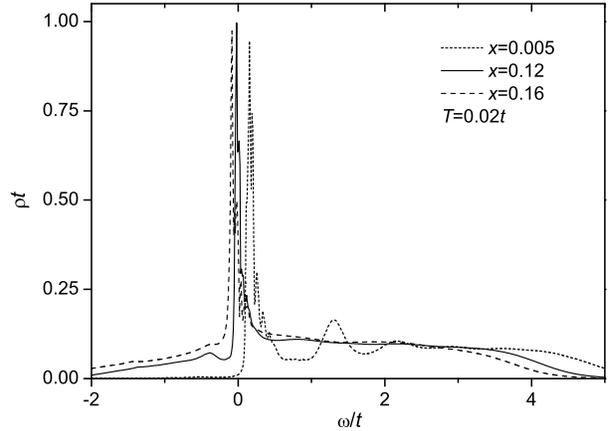}}
\caption{\label{Fig_v}The normalized hole density of states.}
\end{figure}
The normalized hole density of states,
$$\rho(\omega)=N^{-1}\sum_{\bf k}A({\bf k}\omega),$$
is shown in Fig.~\ref{Fig_v} for three hole concentrations from the
extremely low to the moderate doping case. As seen from the figure, a
considerable part of the density of states is concentrated at a maximum
which persists near the Fermi level in a wide range of hole
concentrations. This maximum is produced by the above-mentioned
extended van Hove singularities in the spin-polaron band and plays a
great role in the superconductive transition in the $t$-$J$ model
\cite{Sherman95}.

\section{The spectrum of spin excitations}
\begin{figure}
\centerline{\includegraphics[width=7cm]{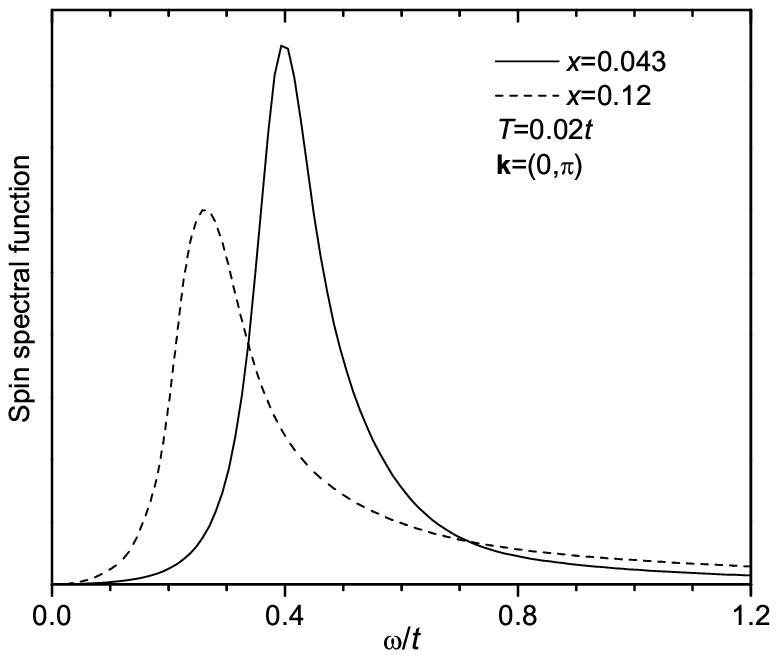}}

\vspace{4mm}\centerline{\includegraphics[width=7.05cm]{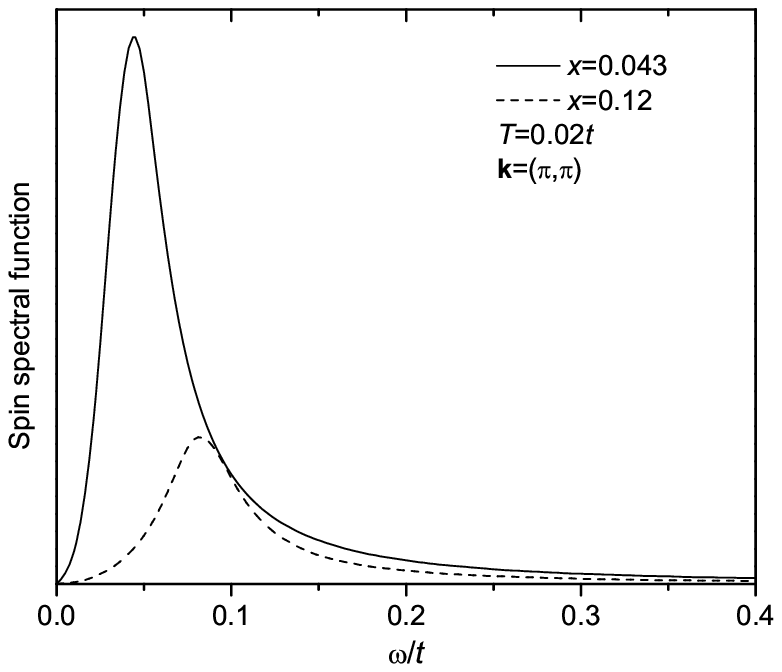}}
\caption{\label{Fig_vi}The spin spectral functions for ${\bf
k}=(0,\pi)$ and $(\pi,\pi)$.}
\end{figure}

Typical shapes of the spin spectral function $B({\bf k}\omega)$
are shown in Fig.~\ref{Fig_vi}.
For low and moderate $x$ the spectrum consists of a maximum with an
extended high-frequency tail. Some fine structure can be observed for
certain wave vectors which is connected with the frequency dependence
of the polarization operator (\ref{po}).

The increases of the temperature and the hole concentration act in a
similar manner on the spectrum. This action is different in the nearest
vicinity of $(\pi,\pi)$ and in the remainder of the Brillouin zone. As
seen from Fig.~\ref{Fig_vi}, in the latter case with increasing $x$
(and $T$) the maximum in $B({\bf k}\omega)$ is shifted to lower
frequencies and loses its intensity, while in the former case the
frequency of the maximum, on the contrary, grows. This frequency growth
is connected with the gap in the spin excitation spectrum at
$(\pi,\pi)$. The magnitude of this gap grows with increasing $x$ and
$T$.

\begin{figure}
\centerline{\includegraphics[width=7cm]{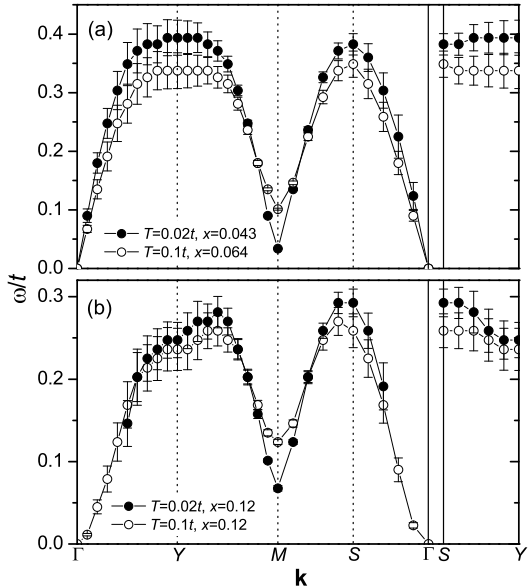}}
\caption{\label{Fig_vii}The dispersion of spin excitations. Vertical
bars show decay widths $|{\rm Im}\Pi({\bf k}\omega)|/(2\omega_{\bf
k})$. The notations of the symmetry points are the same as in
Fig.~\protect\ref{Fig_iii}.}
\end{figure}

As follows from Eq.~(\ref{se}), the frequencies of spin excitations
satisfy the equation
\begin{equation}
\omega^2-{\rm Re}\Pi({\bf k}\omega)-\omega^2_{\bf k}=0. \label{sef}
\end{equation}
Their dispersion along the symmetry lines is shown in
Fig.~\ref{Fig_vii} for different $x$ and $T$. In this figure vertical
bars depict the decay widths $|{\rm Im}\Pi({\bf
k}\omega)|/(2\omega_{\bf k})$ of the excitations, the imaginary parts
of the respective poles in Green's function.
As seen from Fig.~\ref{Fig_vii}a, for low $x$ and $T$ the
dispersion of spin excitations is close to the dispersion of spin
waves. The main difference is the above-mentioned spin gap at
$(\pi,\pi)$. In an infinite crystal the magnitude of this gap is
directly connected with the spin correlation length $\xi$. Indeed,
using Eq.~(\ref{se}) and taking into account that the region near
$(\pi,\pi)$ gives the main contribution to the summation over {\bf
k}, we find for large distances and low temperatures
\begin{eqnarray}
\left\langle s^z_{\bf l}s^z_{\bf 0}\right\rangle&=&N^{-1}\sum_{\bf
k}{\rm e}^{i\bf kl}\int_0^\infty
d\omega\coth\left(\frac{\omega}{2T}\right) B({\bf k}\omega)\nonumber\\
&\propto&{\rm e}^{i\bf Ql}(\xi/|{\bf l}|)^{1/2}{\rm e}^{-|{\bf
l}|/\xi},\label{spincor}\\
\xi&=&\frac{a}{2\sqrt{\Delta}},\nonumber
\end{eqnarray}
where ${\bf Q}=(\pi,\pi)$ and $a$ is the intersite distance (in the
considered temperature and hole concentration ranges $\omega_{\bf Q}^2$
$\gg |{\rm Re}\, \Pi({\bf Q},\omega_{\bf Q})|$; therefore in the above
equation the gap magnitude is approximated by $\omega_{\bf Q}$). For
low $x$ we found $\Delta\approx 0.2x$ and consequently
$$\xi\approx \frac{a}{\sqrt{x}}.$$
This relation between the spin correlation length and the hole
concentration has been experimentally observed in
La$_{2-x}$Sr$_x$CuO$_4$ \cite{Keimer}.

\begin{figure}
\centerline{\includegraphics[width=8.5cm]{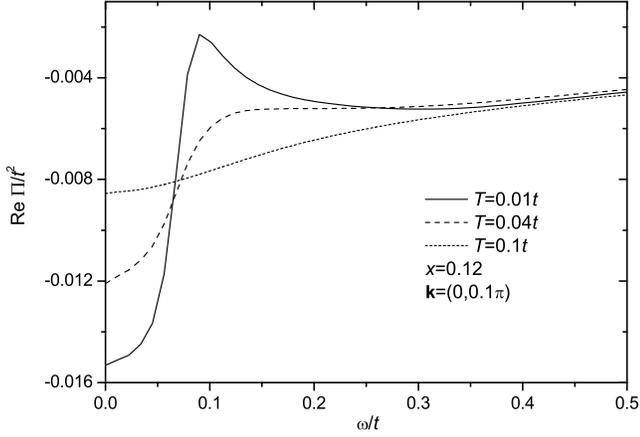}}
\caption{\label{Fig_viii} The real part of the polarization operator
${\rm Re}\, \Pi({\bf k}\omega)$ for ${\bf k}=(0,0.1\pi)$, $x=0.12$ and
various temperatures.}
\end{figure}

With growing $x$ the spin excitation branch is destroyed in some region
around the $\Gamma$ point -- for such momenta Eq.~(\ref{sef}) has no
real solution due to a negative value of ${\rm Re} \, \Pi({\bf
k}\omega)$. In Fig.~\ref{Fig_vii} this peculiarity is reflected in the
rupture of the dispersion branch in this region where excitation
frequencies are purely imaginary. Thus, much like the Heisenberg model
\cite{Tserkovnikov,Reger}, properties of elementary excitations near
$(0,0)$ and $(\pi,\pi)$ are different: in the former region the
excitations are overdamped, while in the latter they are gapped. With
rise of temperature the branch are partly restored around $(0,0)$. The
explanation for such behavior follows from Fig.~\ref{Fig_viii}. For low
temperatures and long wavelengths ${\rm Re} \, \Pi({\bf k}\omega)$ has
a pronounced dip at low frequencies. This dip is connected with the
spin-polaron band and is responsible for the lack of real solutions in
Eq.~(\ref{sef}). With increasing $T$ the spin-polaron peaks are
smeared, the depth of the dip in Fig.~\ref{Fig_viii} becomes smaller
and Eq.~(\ref{sef}) has again real solutions.

\section{Magnetic properties}
As already mentioned, in accord with the Mermin-Wagner theorem
\cite{Mermin} the considered 2D system is in the paramagnetic state for
$T > 0$. This result can also be obtained using equations of Sec.~II\@.
Numerical calculations and the analysis of experimental data presented
strong evidence that the 2D nearest-neighbor $s=\frac{1}{2}$ Heisenberg
antiferromagnet has long-range order at $T=0$ \cite{Reger}. Let us
consider an infinite crystal at $T=0$ and find the hole concentration
$x_c$ which destroys this ordering. Notice that the indication of the
long-range order is a finite value of the condensation parameter $C$
which is determined from constraint (\ref{zsm}) and the condition
$\Delta=0$,
\begin{equation}
C=\frac{1}{2}(1-x)-\sqrt{\frac{|C_1|}{\alpha}}\frac{1}{N}\sum_{\bf k
\neq Q}\sqrt{\frac{1-\gamma_{\bf k}}{1+\gamma_{\bf k}}}, \quad N
\rightarrow \infty. \label{condparam}
\end{equation}
As seen from Eqs.~(\ref{zsm}) and (\ref{condparam}),
$$C=\sqrt{\frac{2|C_1|}{\alpha}}\lim_{N \rightarrow
\infty}\frac{1}{N\sqrt{\Delta}}.$$ As follows from Eq.~(\ref{spincor}),
the sublattice magnetization $M=(\lim_{|{\bf l}| \rightarrow \infty}
|\langle{\bf s_l \cdot s_0} \rangle|)^{1/2}$ is connected with the
condensation parameter by the relation $M=(3C/2)^{1/2}$. To estimate
$x_c$ the dependencies of $C_1$ and $\alpha$ on $x$ have to be taken
into account. From our calculations for low $x$ we found that $|C_1|
\approx 0.2117-0.5750x$ and $\alpha \approx 1.802-8.021x$. Substituting
these values into Eq.~(\ref{condparam}) we find that $C$ vanishes at
$x_c \approx 0.02$. Thus, at $T=0$ the infinite 2D crystal is in the
state with long-range antiferromagnetic order for $x < x_c$ and in the
paramagnetic state for $x > x_c$.

\begin{figure}
\centerline{\includegraphics[width=8.cm]{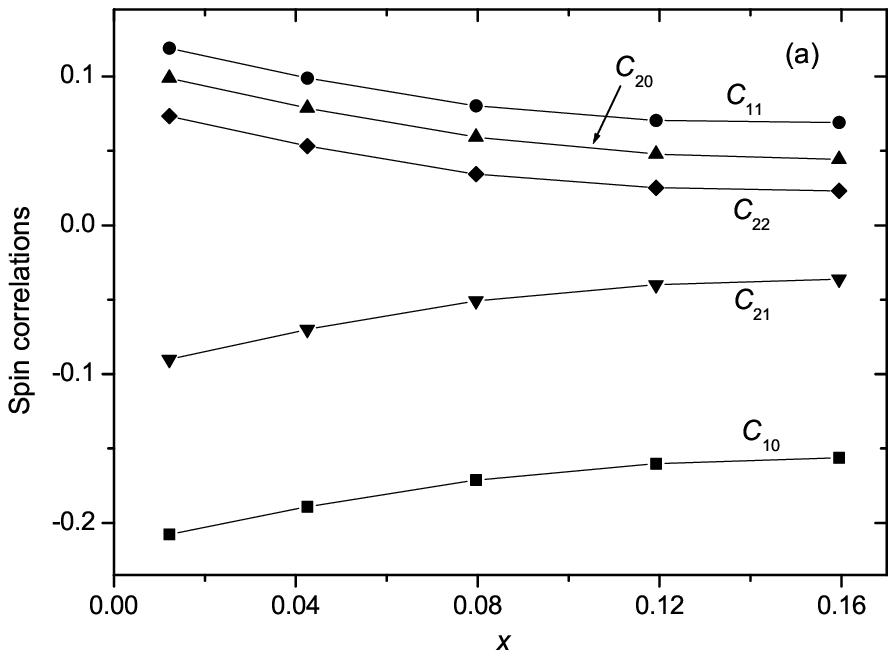}}

\vspace{2mm} \centerline{\includegraphics[width=8.cm]{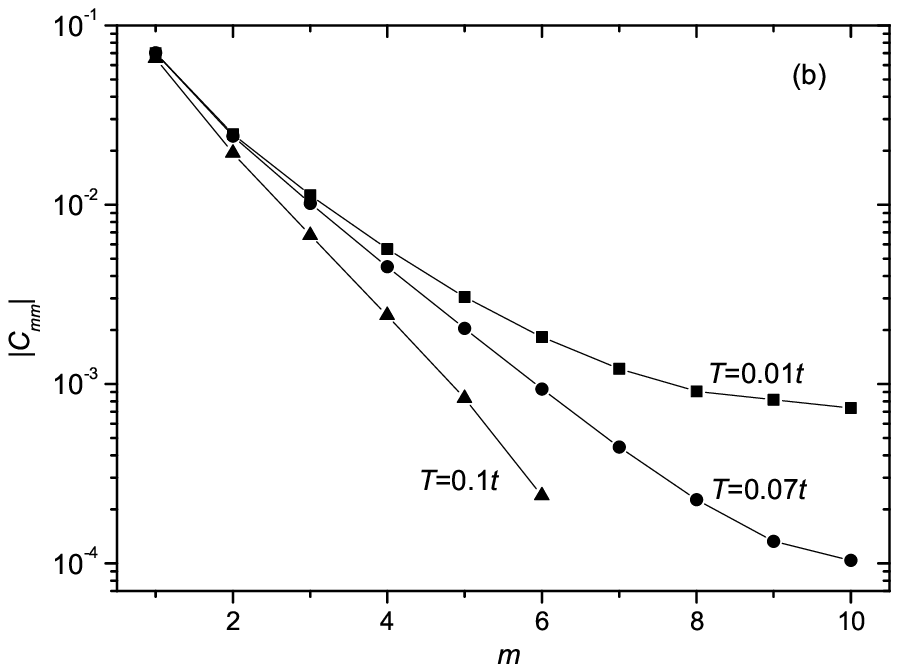}}
\caption{\label{Fig_ix} (a) Spin correlations vs.\ $x$ for $T=0.02t$.
(b) Spin correlations along the diagonal of the crystal [i.e.\ ${\bf
l}=(m,m)$] for $x=0.12$. The respective temperatures are indicated near
the curves.}
\end{figure}

In the considered finite lattice the correlation length is limited by
the size of the lattice. The spin correlations $C_{mn}=\langle s^z_{\bf
l}s^z_{\bf 0}\rangle$, ${\bf l}=(m,n)$ calculated from the obtained
spin spectral function with the use of Eq.~(\ref{spincor}) are shown in
Fig.~\ref{Fig_ix}.
For large enough $x$ and $T$ the correlations decay exponentially with
distance in the considered finite lattice. As mentioned, the method
used has no preset magnetic ordering. The character of the ordering is
determined in the course of the self-consistent calculations. The
magnetic susceptibility obtained in these calculations is strongly
peaked at the antiferromagnetic wave vector for low frequencies. No
indications of incommensurability, which can be related to stripes or
other types of phase separation, are observed in the susceptibility.
Conceivably such phase separations are not connected with the strong
electron correlations described by the $t$-$J$ model.

\begin{figure}
\centerline{\includegraphics[width=6cm]{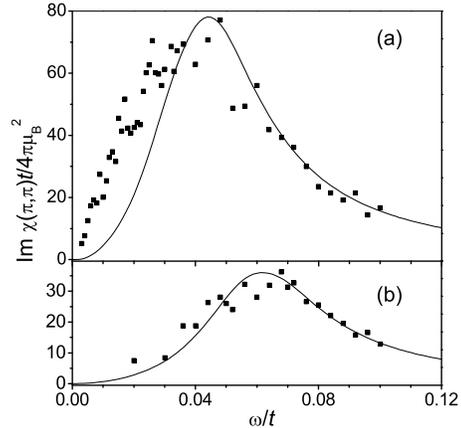}}
\caption{\label{Fig_x}The imaginary part of the spin susceptibility for
${\bf k}=(\pi,\pi)$. Curves show calculated results for $T=0.02t\approx
116$~K, $x=0.043$ (a) and 0.08 (b). Squares are experimental results
obtained in normal-state YBa$_2$Cu$_3$O$_{7-y}$ at $T=100$~K for
$y=0.5$ (a) and 0.17 (b) \protect\cite{Bourges}.}
\end{figure}

The magnetic susceptibility is connected with the spin Green's function
(\ref{se}) by the relation
$$\chi^z({\bf k}\omega)=-4\mu_B^2 D({\bf k}\omega),$$ where $\mu_B$ is the
Bohr magneton. Experiments on inelastic neutron scattering give
information on the susceptibility which can be directly compared with
the calculated results. Such comparison with the results measured in
normal-state YBa$_2$Cu$_3$O$_{7-y}$ \cite{Bourges} is carried out in
Fig.~\ref{Fig_x}.
YBa$_2$Cu$_3$O$_{7-y}$ is a bilayer crystal and the symmetry allows one
to divide the susceptibility into odd and even parts. For the
antiferromagnetic intrabilayer coupling the odd part can be compared
with the calculated results. The oxygen deficiencies $y=0.5$ and 0.17
in the experimental data in Fig.~\ref{Fig_x} correspond to the hole
concentrations $x \approx 0.05$ and 0.11, respectively \cite{Tallon}.
As seen from Fig.~\ref{Fig_x}, the calculated data reproduce correctly
the frequency dependence of the susceptibility, the values of the
frequency for which ${\rm Im}\chi(\pi,\pi)$ reaches maximum and their
evolution with doping. The increase of the frequency of the maximum
with $x$ reflects the respective growth of the spin gap (see
Fig.~\ref{Fig_vi}). In absolute units the calculated maxima of ${\rm
Im}\chi(\pi,\pi)$ are $1.5-2$ times larger than the experimental values
which is connected with some difference in decay widths of spin
excitations.

\begin{figure}
\centerline{\includegraphics[width=8.cm]{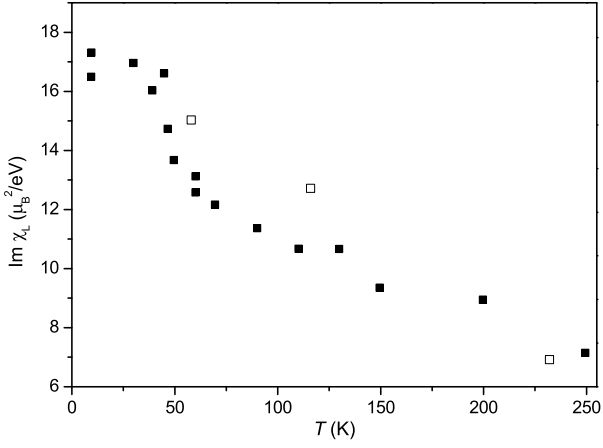}}
\caption{\label{Fig_xi} The local spin susceptibility. The experimental
results \protect\cite{Bourges} in YBa$_2$Cu$_3$O$_{6.5}$ for
$\omega=25$~meV are shown by filled squares. Our calculated data for
$x=0.05$ and $\omega=22.5$~meV are displayed by open squares.}
\end{figure}

The temperature variations of the experimental and calculated
susceptibilities are compared in Fig.~\ref{Fig_xi}.
This figure demonstrates the imaginary part of the local spin
susceptibility which is defined as
$${\rm Im}\chi_L(\omega)=N^{-1}\sum_{\bf k}{\rm Im}\chi({\bf
k}\omega).$$ As seen from the figure, the calculated temperature
variation of the susceptibility is also in good agreement with
experiment.

\begin{figure}
\centerline{\includegraphics[width=8.cm]{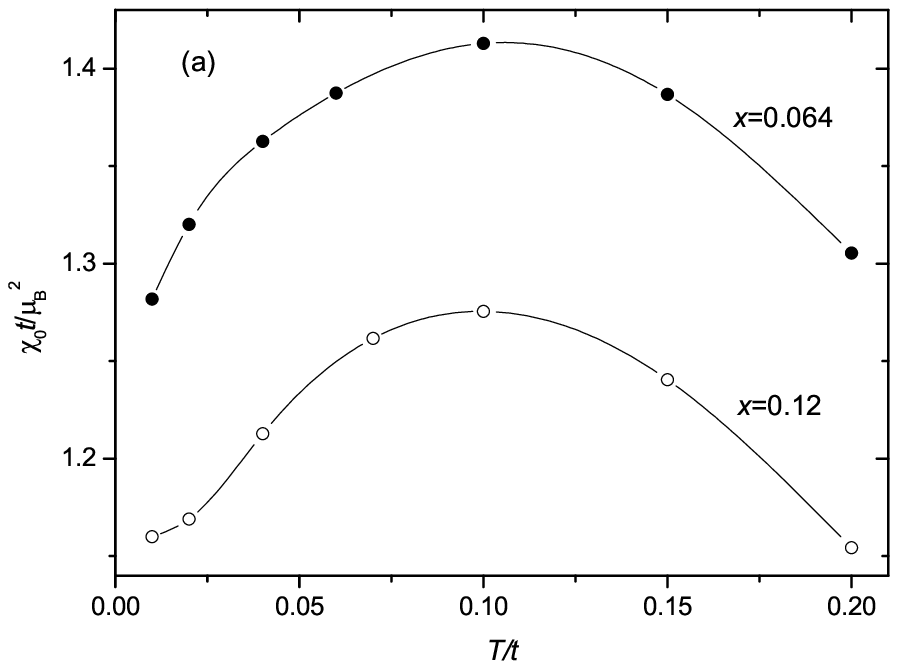}}

\vspace{4mm}\centerline{\includegraphics[width=8.cm]{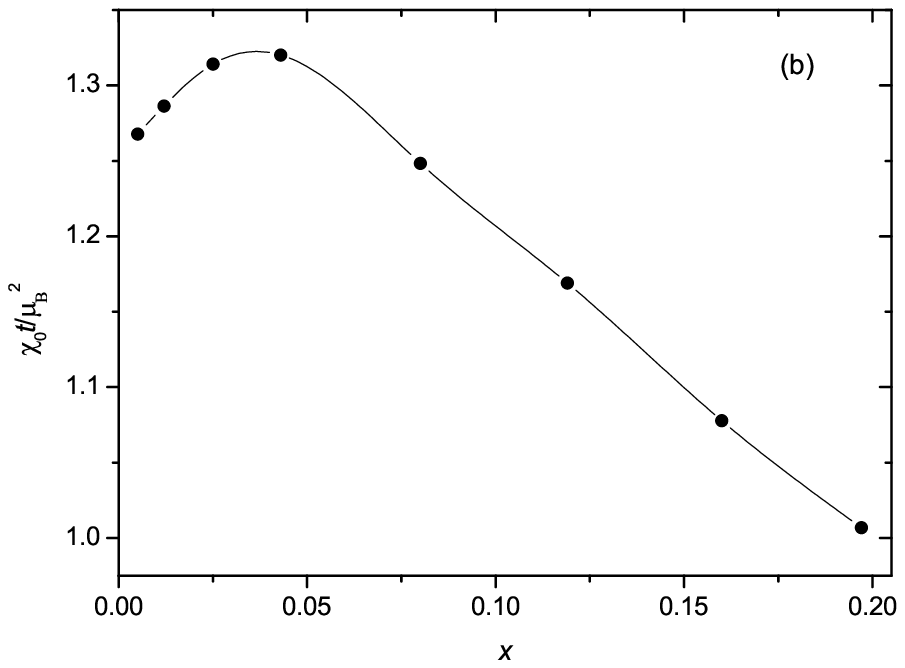}}
\caption{\label{Fig_xii} The uniform static spin susceptibility vs.\
temperature (a) and hole concentration for $T=0.02t$ (b).}
\end{figure}

The calculated temperature and concentration dependencies of the
uniform static spin susceptibility,
$$\chi_0=\chi({\bf k}\rightarrow 0,\omega=0)=4\mu_B^2T^{-1}\sum_{\bf
n}\left\langle s^z_{\bf n}s^z_{\bf 0}\right\rangle,$$ are shown in
Fig.~\ref{Fig_xii}.
The values lie in the range 2--2.6~eV$^{-1}$ which is close to the
values 1.9--2.6~eV$^{-1}$ obtained for YBa$_2$Cu$_3$O$_{7-y}$
\cite{Monien}. Close values of $\chi_0$ were also obtained by the exact
diagonalization of small clusters \cite{Jaklic}. The temperature
dependence of $\chi_0$ has a maximum and the temperature of the maximum
$T_m$ grows with decreasing $x$. Analogous behavior is observed in
cuprates for large enough $x$
\cite{Johnston,Torrance,Takigawa91,Nakano}. In Fig.~\ref{Fig_xii}a $T_m
\approx 600$~K which is close to the value observed in
La$_{2-x}$Sr$_x$CuO$_4$ for comparable hole concentrations
\cite{Johnston}. As known, in the undoped antiferromagnet $T_m \approx
J$ \cite{Manousakis}. On the high-temperature side $\chi_0(T)$ tends to
the Curie-Weiss behavior $1/T$.

The decrease of $\chi_0$ below $T_m$ is sometimes considered as the
manifestation of the spin gap. In our opinion this statement is
incorrect. For moderate $x$ and $T$ the long-wavelength part of the
spin excitation spectrum does not feel the gap at $(\pi,\pi)$. For
small but finite values of {\bf k} $\chi({\bf k},0)\propto
\int_{-\infty}^\infty d\omega' B({\bf k}\omega')/\omega'$. As indicated
in Sec.~IV, the function $B({\bf k}\omega')$ has a maximum which is
shifted to lower frequencies and loses its intensity with increasing
temperature for such wave vectors. In the above integral the maximum is
superimposed with the decreasing function $1/\omega'$ which finally
leads to the nonmonotonic behavior of $\chi_0(T)$.

As seen from Fig.~\ref{Fig_xii}a, the two curves for the different $x$
are very close in shape and can be superposed by scaling to the same
values of the maximum $\chi_0$ and $T_m$. Analogous scaling was
observed in La$_{2-x}$Sr$_x$CuO$_4$ \cite{Johnston}. As follows from
the above discussion, the source of this scaling is that holes and
temperature fluctuations lead in a similar manner to the softening of
the maximum in $B({\bf k}\omega')$ for long wavelengths.

As in experiment \cite{Johnston,Torrance}, the dependence $\chi_0(x)$
in Fig.~\ref{Fig_xii}b has a maximum. However, the calculated value of
the hole concentration which corresponds to the maximum is much smaller
than the experimental value $x\approx 0.25$. A possible reason for this
discrepancy is the approximation used for calculating the hole
self-energy \cite{Sherman02}.

The spin-lattice relaxation and spin-echo decay rates were calculated
with the use of the equations \cite{Barzykin}
\begin{eqnarray}
\frac{1}{^\alpha T_{1\beta}\,T}&=&\frac{1}{2\mu^2_BN}\sum_{\bf k}
 \,^\alpha\!
 F_\beta({\bf k})\frac{{\rm Im}\,\chi({\bf k}\omega)}{\omega},
 \quad\omega\rightarrow 0,\nonumber\\
\frac{1}{({^{63}T_{2G}})^2}&=&\frac{0.69}{128\mu_B^4}\Biggl\{
 \frac{1}{N}\sum_{\bf k}\,^{63}\!F_e^2({\bf k})\left[{\rm Re}\,
 \chi({\bf k}0)\right]^2 \label{nmr}\\
&-&\left[\frac{1}{N}\sum_{\bf k}\,^{63}\!F_e({\bf k})
 {\rm Re}\,\chi({\bf k}0)\right]^2\Biggr\}, \nonumber
\end{eqnarray}
where the form factors are
\begin{eqnarray}
&&^{63}\!F_{\|}({\bf k})=\left(A_\bot+4B\gamma_{\bf
k}\right)^2,\nonumber\\
&&^{63}\!F_e({\bf k})=\left(A_\|+4B\gamma_{\bf k}\right)^2,\nonumber\\
&&\label{formfactors}\\
&&^{63}\!F_\bot({\bf k})=\frac{1}{2}\left[^{63}\!F_{\|}({\bf k})+
 ^{63}\!\!F_e({\bf k})\right],\nonumber\\
&&^{17}\!F_\|({\bf k})=2C^2\left(1+\gamma_{\bf k}\right).\nonumber
\end{eqnarray}
In the above formulas the hyperfine coupling constants $B=3.82 \cdot
10^{-7}$~eV, $A_\bot=0.84B$, $A_\|=-4B$, and $C=0.91B$ \cite{Barzykin}.
The superscripts $\alpha=63$ or 17 indicate that the respective
quantity belongs to the Cu or O site, respectively. The subscripts $\|$
and $\bot$ refer to the direction of the applied static magnetic field
{\bf H} with respect to the axis {\bf c} perpendicular to the Cu-O
plane. The form factor $^{63}\!F_e$ is the filter for the Cu spin-echo
decay time $^{63}T_{2G}$. Due to the different momentum dependencies of
the form factors (\ref{formfactors}) measurements of the spin-lattice
and spin-echo decay rates allow one to extract the information on the
low-frequency susceptibility in different regions of the Brillouin
zone.

\begin{figure}
\centerline{\includegraphics[width=8.5cm]{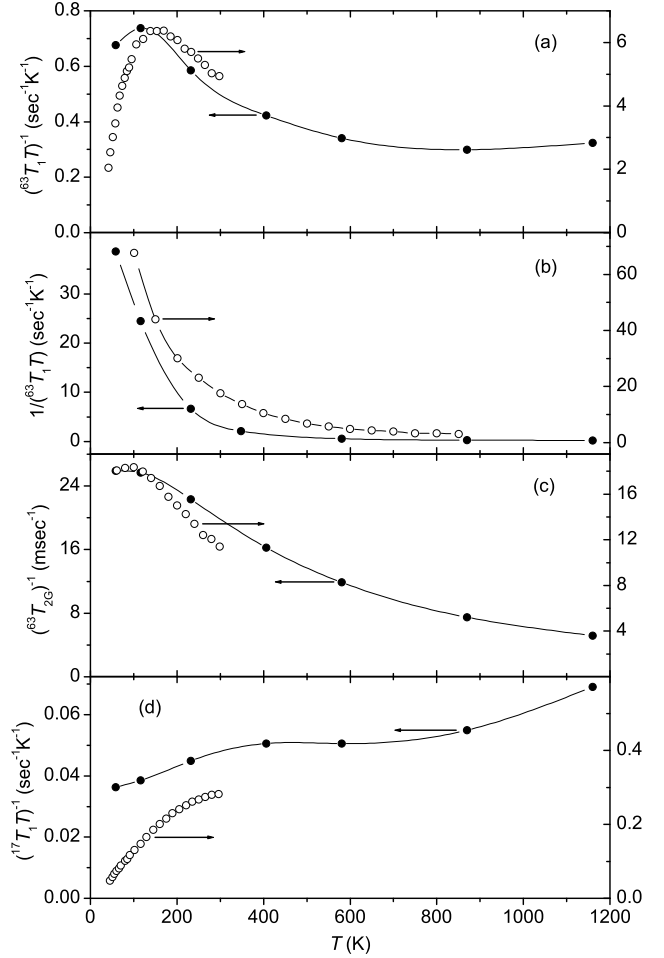}}
\caption{\label{Fig_xiii}The temperature dependencies of the
spin-lattice relaxation and spin-echo decay rates at Cu (a--c) and O
(d) sites. Open circles with right axes represent experimental results,
filled circles with left axes are our calculations. (a,c,d)
Calculations for ${\bf H\|c}$ and $x=0.12$, measurements
\protect\cite{Takigawa91,Takigawa94} in YBa$_2$Cu$_3$O$_{6.63}$
($x\approx 0.1$ \protect\cite{Tallon}). (b) Calculations for
nonoriented configuration with $x=0.043$, measurements
\protect\cite{Imai} in La$_{1.96}$Sr$_{0.04}$CuO$_4$.}
\end{figure}

Our calculated results are compared with the respective experimental
data in Fig.~\ref{Fig_xiii}.
The calculations reproduce satisfactorily the main peculiarities of the
temperature dependencies of the spin-lattice relaxation and spin-echo
decay rates. The growth of $(^{63}T_1\,T)^{-1}$ with decreasing $x$ is
connected with the increase of the spectral intensity of spin
excitations near $(\pi,\pi)$ which make the main contribution to this
rate. For the same hole concentration $(^{63}T_1\,T)^{-1}$ is one order
of magnitude larger than $(^{17}T_1\,T)^{-1}$. This is a consequence of
the fact that ${\rm Im\,}\chi$ is strongly peaked near $(\pi,\pi)$ and
the momentum dependencies of the form factors (\ref{formfactors})
\cite{Barzykin}. The calculated spin-lattice relaxation rates are
smaller than the experimental values due to the approximation made in
the calculation of $D({\bf k}\omega)$ which somewhat underestimates
${\rm Im\,}\chi$ at low frequencies.

\begin{figure}
\centerline{\includegraphics[width=8cm]{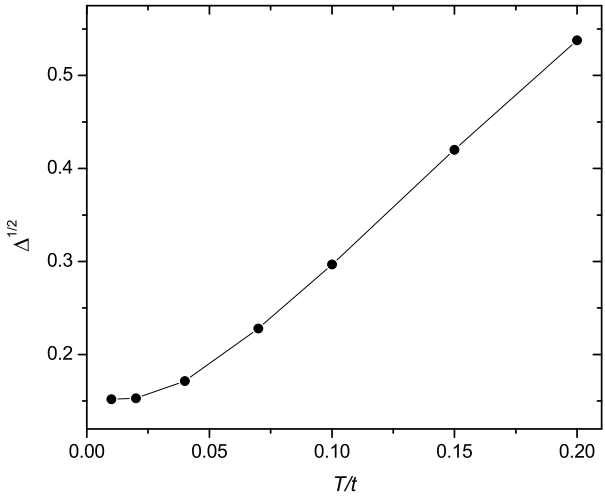}}
\caption{\label{Fig_xiv}The temperature dependency of $\sqrt{\Delta}
\propto \xi^{-1}$ for $x=0.12$.}
\end{figure}

For moderate $x$ with increasing $T$ the low-frequency region of
${\rm Im}\,\chi({\bf k\approx Q})$, ${\bf Q}=(\pi,\pi)$ first
grows due to the temperature broadening of the maximum in its
frequency dependence and then decreases due to the temperature
growth of the spin gap (see Fig.~\ref{Fig_vi} and the related
discussion). This nonmonotonic behavior of the susceptibility
shows up in the spin-lattice relaxation rate at Cu in
Fig.~\ref{Fig_xiii}a. The temperature variations of ${\rm
Im}\,\chi$ and $(^{63}T_1\,T)^{-1}$ can be related with the
temperature behavior of the magnetic correlation length. As can be
seen in Fig.~\ref{Fig_xiv}, for moderate $x$ and low $T$ the
magnitude of the spin gap is determined by the hole concentration
and does not depend on $T$.
This temperature range corresponds to the growth stage in
Fig.~\ref{Fig_xiii}a. As follows from Eq.~(\ref{spincor}), the
independence of the gap from $T$ means that in this temperature range
$\xi$ does not depend on temperature either which is a distinctive
feature of the quantum disordered regime \cite{Barzykin,Chubukov}. For
temperatures above the maximum of $(^{63}T_1\,T)^{-1}$ we found that
$^{63}T_1\,T/^{63}T_{2G}\approx$ const (see Figs.~\ref{Fig_xiii}a and
c). As seen in Fig.~\ref{Fig_xiv}, in this temperature range
$\xi^{-1}\propto\sqrt{\Delta}$ varies linearly with $T$. Both these
facts indicate that in the mentioned temperature range the crystal is
in the quantum critical $z=1$ regime. These conclusions are consistent
with the phenomenological treatment of experiment in
YBa$_2$Cu$_3$O$_{7-y}$ carried out in Ref.~\cite{Barzykin}. The
temperature of the maximum in Fig.~\ref{Fig_xiii}a is close to the
parameter $T_*$ of that work, the temperature which separates the
quantum disordered and quantum critical $z=1$ regimes.

For small hole concentrations the temperature range in which holes
determine the correlation length is absent or very small. In these
conditions the spin gap grows starting from low temperatures and
$(^{63}T_1\,T)^{-1}$ decreases monotonously, as shown in
Fig.~\ref{Fig_xiii}b.

Due to the form factor $^{17}\!F_\|({\bf k})$, Eq.~(\ref{formfactors}),
the momentum region near $(\pi,\pi)$ does not contribute to
$(^{17}T_1\,T)^{-1}$. As indicated in Sec.~IV, there is a cardinal
difference between the behavior of ${\rm Im}\,\chi$ for ${\bf k\approx
Q}$ and away from $(\pi,\pi)$. Due to the spin gap in the former case
the frequency of the maximum in ${\rm Im}\,\chi(\omega)$ increases with
temperature, while in the latter case it decreases. This frequency
softening leads to the growth of the low-frequency ${\rm Im}\,\chi$ and
$(^{17}T_1\,T)^{-1}$ at low $T$ and their saturation for higher
temperatures. Analogous behavior is observed in experiment
\cite{Takigawa94}, as seen in Fig.~\ref{Fig_xiii}d. Thus, the
temperature and concentration variations of the spin excitation
spectrum in the $t$-$J$ model put forward the simple explanations for
the behavior of the spin-lattice relaxation and spin-echo decay rates
observed in cuprates.

\section{Concluding remarks}
In this work we applied Mori's projection operator technique and
the decoupling of the many-particle Green's functions for
obtaining the closed set of self-energy equations which describes
energy and magnetic properties of the $t$-$J$ model of the Cu-O
planes in perovskite high-$T_c$ superconductors. The equations
retain the rotation symmetry of spin components and zero site
magnetization in the paramagnetic state which is set for $T>0$ or
$x > x_c \approx 0.02$ in the case $T=0$ in an infinite crystal.

For the parameters of cuprates in the cases of low and moderate
doping this self-consistently calculated solution of the equation
is homogeneous. This result indicates that phase separations are
not related to strong electron correlations described by the
model.

A number of unusual spectral and magnetic properties of cuprate
perovskites is satisfactorily reproduced by the calculations.
Among these properties are the extended van Hove singularities
around $(0,\pi)$ and $(\pi,0)$. Due to strong electron
correlations these singularities persist near the Fermi level in a
wide range of hole concentrations which has to play an essential
role in the superconductivity of cuprates.

The pseudogap in the calculated hole spectrum has the same
symmetry and is close in magnitude to the pseudogap observed in
photoemission of these crystals. As in experiment, the calculated
pseudogap disappears when the hole concentration approaches the
optimal doping. The existence of two maxima with similar
dispersions in the photoemission spectrum of lightly doped
Ca$_2$CuO$_2$Cl$_2$ is related to hole vibronic states in the
region of the disturbed antiferromagnetic order.

In the considered model the concentration dependence of the
magnetic correlation length is the same as that observed in
La$_{2-x}$Sr$_x$CuO$_4$. As indicated, in contrast to this
crystal, the model does not show low-frequency incommensurate spin
fluctuations for the considered parameters.

In other respects the calculated magnetic susceptibility is close
to that derived from experiments. The results of the calculations
offer explanations for the observed scaling of the static uniform
susceptibility and for the changes in the spin-lattice relaxation
and spin-echo decay rates in terms of the temperature and doping
variations in the spin excitation spectrum. The scaling is related
to the fact that holes and temperature fluctuations lead in a
similar manner to the softening of the maxima in the spin spectral
function for long wavelengths. This softening leads to the
monotonic increase of the spin-lattice relaxation rate at O.

Contrastingly, due to the spin gap the frequencies of the maxima in the
spin spectral function for momenta near $(\pi,\pi)$ grow with
temperature and hole concentration. This region of momenta makes the
main contribution to the spin-lattice relaxation rate at Cu. For low
hole concentrations the growth of the frequencies of the maxima leads
to the monotonic decrease of the rate. For moderate concentrations two
temperature regions may be distinguished. In the low-temperature region
the gap magnitude and the correlation length depend only weakly on
temperature. In this region the spin-lattice relaxation rate at Cu
grows. The behavior of these parameters in the high-temperature region
is the same as for low hole concentrations.

This work was partially supported by the ESF grant No.~4022 and by DFG.

\end{document}